**Dynamic Structural Impact of the COVID-19 Outbreak on the Stock Market and the Exchange Rate: A Cross-country Analysis Among BRICS Nations**

Rupam Bhattacharyya[*], *Department of Biostatistics, University of Michigan. Address: 1415 Washington Heights, Ann Arbor, MI 48109, USA. Email ID: rupamb@umich.edu, Phone: +17348006834;*

Sheo Rama, *Department of Economic Studies and Policies, Central University of South Bihar, Bihar, India. Email ID: sheorama1810@gmail.com;*

Atul Kumar, *Department of Economic Studies and Policies, Central University of South Bihar, Bihar, India. Email ID: atulsingh.atul960@gmail.com;*

Indrajit Banerjee, *Department of Economic Studies and Policies, Central University of South Bihar, Bihar, India. Email ID: indrajitbanerjee03@gmail.com.*

[*]*Corresponding author.*

# ABSTRACT

COVID-19 has impacted the economy of almost every country in the world. Of particular interest are the responses of the economic indicators of developing nations (such as BRICS) to the COVID-19 shock. As an extension to our earlier work on the dynamic associations of pandemic growth, exchange rate, and stock market indices in the context of India, we look at the same question with respect to the BRICS nations. We use structural variable autoregression (SVAR) to identify the dynamic underlying associations across the normalized growth measurements of the COVID-19 cumulative case, recovery, and death counts, and those of the exchange rate, and stock market indices, using data over 203 days (March 12 – September 30, 2020). Using impulse response analyses, the COVID-19 shock to the growth of exchange rate was seen to persist for around 10+ days, and that for stock exchange was seen to be around 15 days. The models capture the contemporaneous nature of these shocks and the subsequent responses, potentially guiding to inform policy decisions at a national level. Further, causal inference-based analyses would allow us to infer relationships that are stronger than mere associations.

**INTRODUCTION**

With the four fastest growing developing nations - Brazil, Russia, India, and China - an organization termed BRIC in short was first brought together at Yekaterinburg, Russia in 2008 for its first summit, the term having been in use since the beginning of the millennium (O'neill, 2001). In 2010, South Africa joined in, and BRICS was formed (Nkoane-Mashabane, 2012). The BRICS represents 41.42% of the total population in the world and contributes 24.10% of the world's total GDP, as in 2019, with a 16.25% export share (BRICS Policy Center, 2019). BRICS is unique in that it is a successful international organization with developing nations from geographically different parts of the world. The growth rate of population in the member nations of BRICS combinedly decreased by 3.20% from 2010 to 2019 but the percentage share in GDP increases by 34.04% with an average increment of 3.80% yearly during the same period (Appendices 1-2) (The World Bank, 2021). The BRICS countries were no exception in terms of being impacted by the global COVID-19 pandemic in different socio-economic ways, and have been the epicentres of recent research interest in quantifying those effects (Dash et al., 2021; Isheloke, 2020).

In a recent study, we explored the associations among the growth rates of infected COVID cases, exchange rate, and SENSEX, in the context of India (Banerjee et al., 2020). Simple correlation-based analyses indicated that the growth rate of COVID-19 confirmed cases was positively correlated with the growth of the exchange rate of Indian currency, but negatively correlated with that of SENSEX. In a further dynamic analysis based on vector autoregressive models (VAR), the associations were not found to be statistically significant but were suggestive, and a time-varying pattern in the directions of the associations was observed, modulated by the non-pharmaceutical interventions such as lockdowns and unlocks implemented by the government (Banerjee et al., 2020). Another recent study focused on the role of the stock market in the Indian economy and the correlation between the performance of the stock market and economic growth, suggesting that the identification of factors affecting stock market is a necessity for government policymakers to formulate and target decisions guiding the nation towards development (Salameh & Ahmad). Several studies in the post-COVID research scene have attempted to answer the question of dynamic shock response in exchange rates and/or stock markets focusing on different locations such as Japan (Narayan et al., 2020), Indonesia (Syahri & Robiyanto, 2020), and Australia (Narayan et al., 2021). Given the significance of the BRICS in the international context, as discussed in the beginning of this section, these nations too deserve a closer investigation of their economic repercussions to the COVID-19.

As a natural extension of our previous work focused on India, we formulate this study to explore the associations among the pandemic and the economic indicators across the BRICS nations over the span of March-September 2020. From a methods perspective, we replace the VAR models used in our previous study by structural VAR (SVAR) models (Gottschalk, 2001). Added to the joint dynamics of a set of variables represented by a VAR model, the structural form depicts the underlying structural relationships, and offers two additional utilities. First, uncorrelated error terms help in separating out the effects of potential economically unrelated influences in the VAR, and further, the model allows variables to have a contemporaneous (rather than only immediate) impact on other variables (Amisano & Giannini, 2012). Using publicly available COVID-19 and economic data and open-source software packages, as described in the next section, we perform our analyses and attempt to interpret the results and address their implications and limitations over the remainder of this paper.

**DATASETS AND METHODS**

**Data Sources**

COVID-19 daily and cumulative confirmed case, recoveries, and deaths data were obtained from the [COVID-19 Data Repository by the Center for Systems Science and Engineering (CSSE) at Johns Hopkins University](#) (Dong et al., 2020). Necessary corrections for Indian COVID-19 incidence data were performed based on the crowdsourced initiative [COVID-19 India](#) (COVID-19 India, 2021). Historical stock market data for the BRICS nations were obtained from the [Yahoo Finance website](#) (Verizon Media, 2020). Historical exchange rate data for the BRICS nations' primary currencies were obtained from the [Exchange Rates UK data dashboard](#) (Exchange Rates UK, 2020).

**Data Processing Methodology**

Our analyses are based on data starting from March 12, 2020, the first day after COVID-19 was deemed a worldwide pandemic by the World Health Organization (WHO) (Jebril, 2020), since this announcement was potentially one of the initiators of the global market shock due to COVID-19, and we used data until September 30, 2020 ($n = 203$). The alignment and clean-up of the COVID-19 and economic data were performed using R 3.6.1 (R Core Team, 2019). Three COVID-19 variables (cumulative counts of confirmed, deceased, and recovered cases) and two economic variables (exchange rate with respect to US dollar, and stock market index) were used for the analyses. Each variable was converted into day-to-day growth rates first (for a variable $X$, growth rate for day $t$ is computed as $G_t^X = 100 \times \frac{X_t - X_{t-1}}{X_{t-1}}$), and then these growth rates were normalized using the R function *scale* to ensure balanced and reliable model fits and comparable magnitude of the estimates. The five final variables were respectively called GrowthC, GrowthD, GrowthR, GrowthER, and GrowthSV.

**Implementation and Summarization of SVAR Models**

The SVAR analyses were performed using the s*vars* and *vars* packages in R (Lange et al., 2019; Pfaff, 2008). Alongside the five variables already defined, an adjustment variable for the country-specific effects not included in our data was included (categorical with five levels, included as an exogenous variable). Choices of optimal lag using the *VARselect* function were obtained as 2 days (Schwarz Criterion), 7 days (Hannan Quinn Criterion), and 13 days (Akaike Information Criterion and Final Prediction Error). We fit a model with each of these lag choices to identify any variation in the shock response patterns. The final SVAR models were fit using the *id.ngml* function utilizing the non-Gaussian maximum likelihood procedure, since visual inspection of the variables indicated deviation from normality (Lanne et al., 2017). Impulse response plots were created using the *irf* function.

**RESULTS**

Figure 1 indicates that all three COVID growth variables were initially higher among all the BRICS nations except China, which is intuitive since at the beginning of our data, China already had a large cumulative number of cases unlike the others. GrowthER remained erratic throughout the period considered, with South Africa and China experiencing higher scales of fluctuation than the others. On the other hand, the fluctuation in GrowthSV was relatively higher for Brazil and South Africa, while it was relatively milder for China. India initially experienced higher fluctuation, which later got milder.

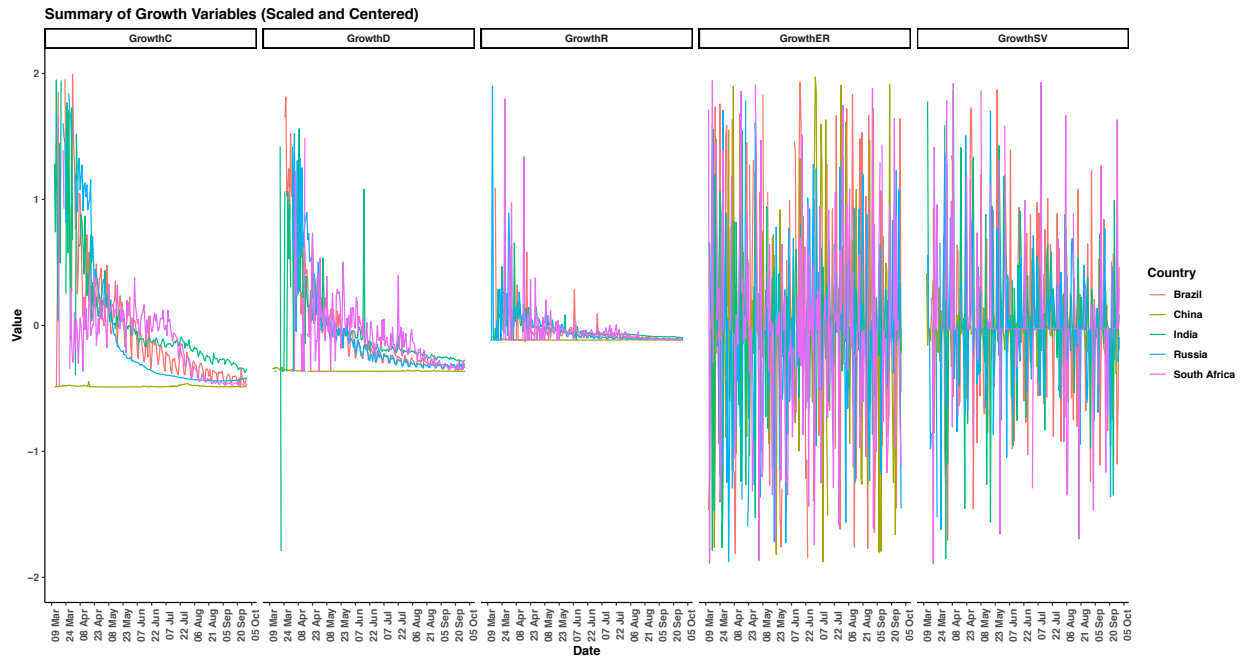

**Figure 1: Time series of normalized growth variables across the BRICS countries over March 12 – September 30, 2020.** The variable acronyms are defined in the methods section.

Figures 2, 3, and 4 exhibit the impulse response functions matrix over 20 forward time units (days) across the five variables for our SVAR model with lags 2, 7, and 13, respectively. In each case, the principal $3 \times 3$ submatrix summarizes the associations within the COVID variables, and the $2 \times 2$ submatrix diagonal to it summarizes those within the economic variables. Since we are primarily interested in the effect of COVID variables on the economic variables, we will focus mostly on the (bottom-left) $2 \times 3$ submatrices summarizing these effects for the rest of this section.

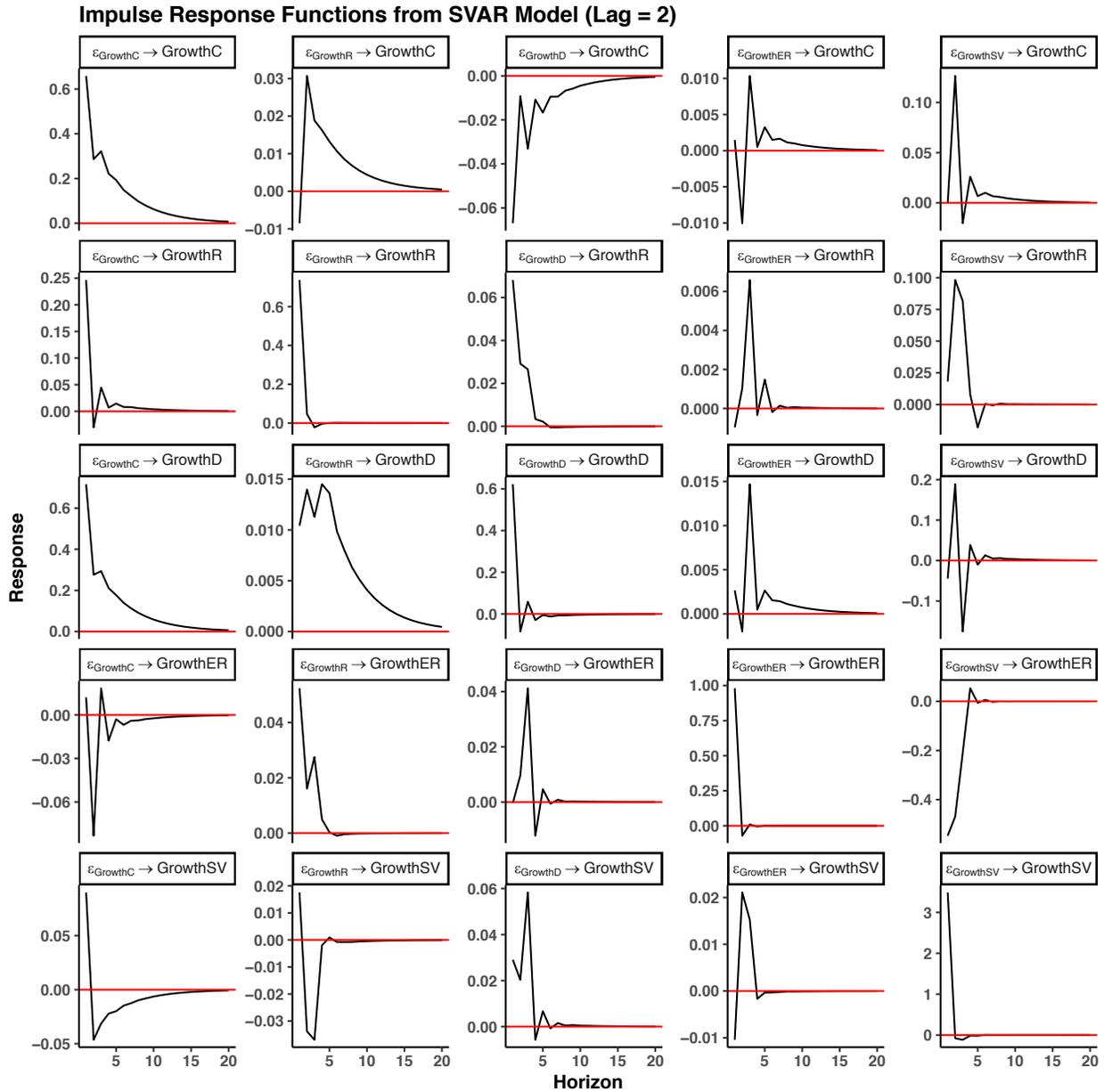

**Figure 2: Impulse response functions for the structural vector autoregressive model with lag 2.** The variable acronyms are defined in the methods section.

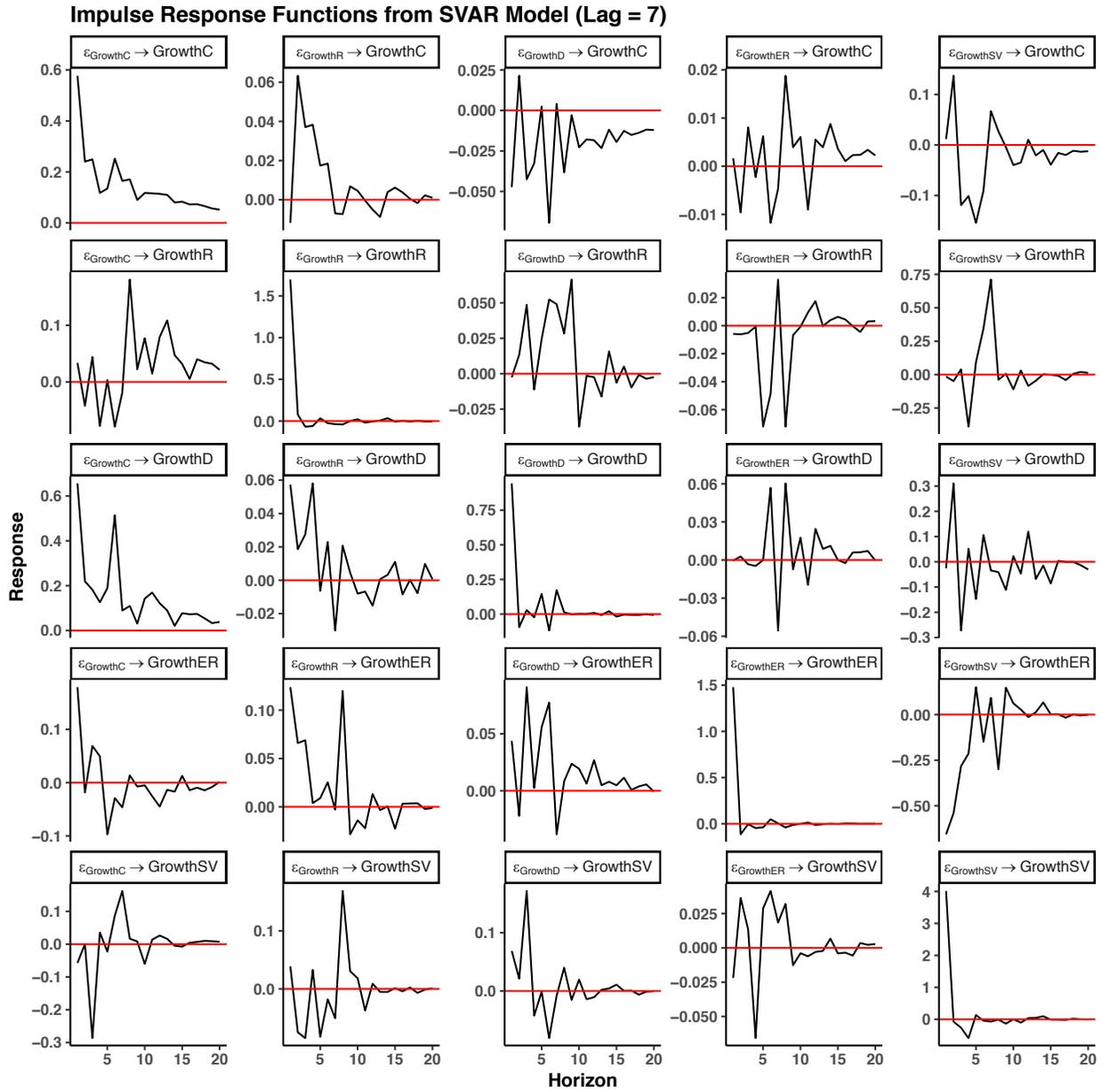

**Figure 3: Impulse response functions for the structural vector autoregressive model with lag 7.** The variable acronyms are defined in the methods section.

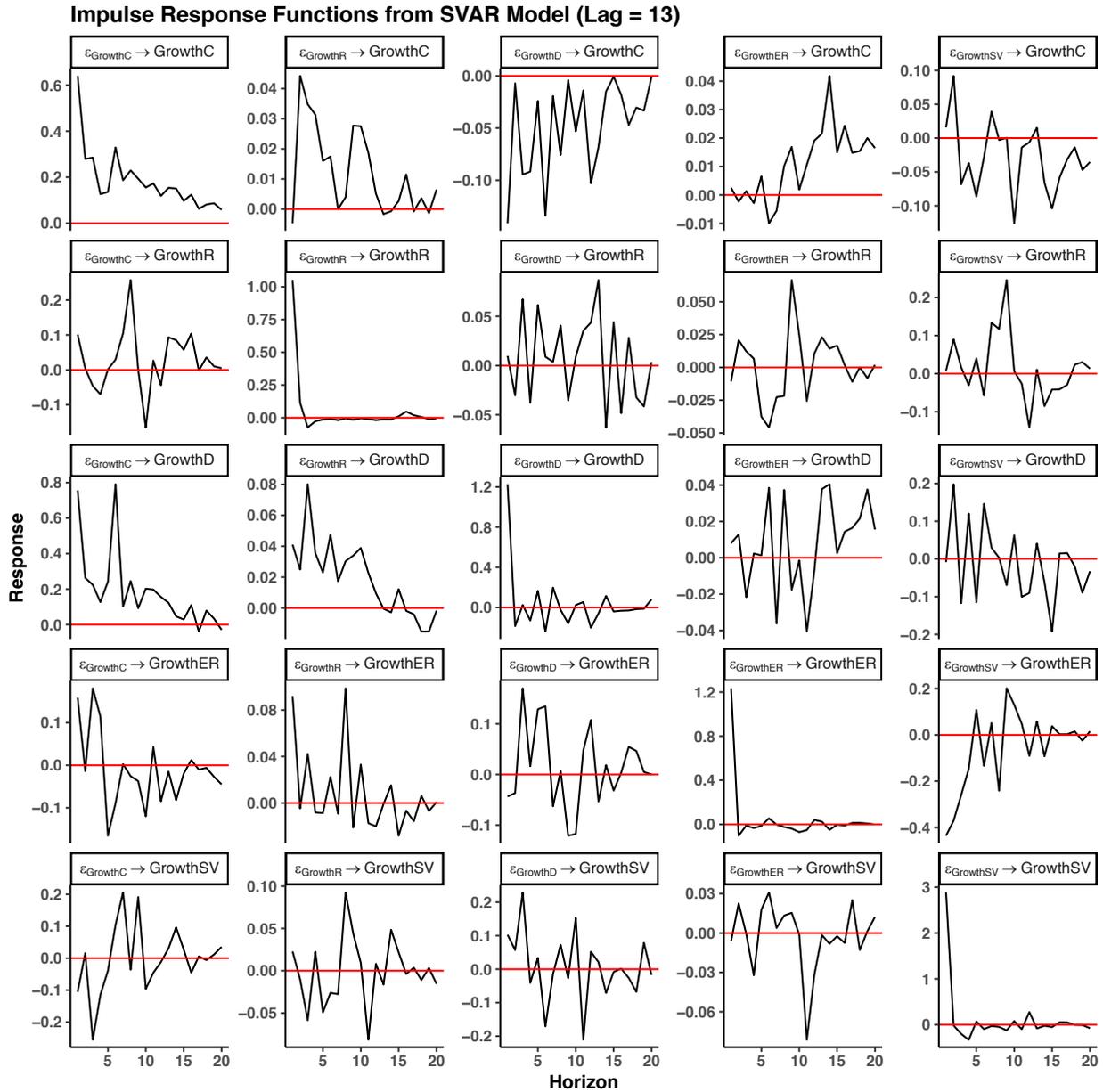

**Figure 4: Impulse response functions for the structural vector autoregressive model with lag 13.** The variable acronyms are defined in the methods section.

Across the three different choices of lag, it can be observed that the absolute length of the deviation from zero due to the shock is more or less similar, and the only feature that differs among these three scenarios are the erratic patterns of the fluctuations captured within that timeframe (Figures

2-4). This is expected since higher lag length allows the model to capture the changes over time at a more granular level. With a few exceptions, the GrowthC and GrowthD variables appear to induce higher absolute magnitudes of shock to the economic variables than those induced by GrowthR (Figures 2-4).

For GrowthER, the shock duration lies somewhere around 10 days across all three models from all three COVID variables, with the post-10 days shocks being very close to zero in magnitude, as captured in Figures 3-4. For GrowthSV, except for the case of lag 2 with GrowthR and GrowthD, the shock duration appears to be relatively larger with stability being reobtained at around 15 days or more, as can be observed in Figures 3-4. We performed further exploratory analyses using possibly higher choices of lag such as 21 and 28 days and visualized the results for higher than 20 units of impulse response horizon (not included here). The resulting estimated response timeframes did not appear to change significantly from those presented here, validating the estimations in these models and indicating that our optimal lag-based models did not tend to underestimate the impulse response durations by too much. We also did not plot the confidence bands associated with these curves since they were quite sharp (as we had $n = 203$ days' worth of data) and would not effectively change the inferences presented here based just on the estimated impulse response curves.

**DISCUSSIONS**

The impact of COVID-19 on economic indicators and the post-COVID economic scenario of the developing nations pose an intriguing set of questions before current researchers, and this paper offers an amalgamation of these two research interests in context of BRICS nations using structural variable autoregression models to identify underlying associations among COVID and economic variables. Our models successfully captured the contemporaneous nature of the COVID shock response of the exchange rate and stock market variables, indicating around one and a half to two weeks of return windows. While direct validation of these results based on recent literature is challenging given that most of the research on COVID-19 in BRICS countries have focused on instantaneous or average effect estimations (Al-Awadhi et al., 2020; Dash et al., 2021) instead of dynamic or time-varying effects, our results align with results from other studies based on the impact of COVID-19 on other economic indicators (e.g. crude oil prices) in other developing and developed parts of the world (Gil-Alana & Monge, 2020; Khurshid & Khan, 2021; Sharif et al., 2020).

The strength of this study stems from several of its methodological features. First of all, our analyses are based on a large sample ($n = 203$ days) and the confidence intervals/bands associated with the estimated coefficients turned out to be extremely precise. The usage of normalized growth variables allows the models to be more stable and the estimates to be less pathological and more reliable. The non-Gaussian maximum likelihood-based model fitting procedure ensures that our results are not impacted by mistaken assumptions, and the four different criteria used for lag estimation ensure a broad coverage of the impact spectrum. The inclusion of the categorical exogenous variable adjusting for country-level variations adjusts for inherent differences among the five economies not otherwise captured in our data. One limitation of our approach is in that the data, although a time series, is observational in nature, and the statistical methodology used only looks for associations among the variables – thereby not allowing us to make causal statements. Some recent literature have been focused on usage of methods that can establish causal links between COVID-19 and socio-economic variables, and application of such methods in context of BRICS nations offer a potential window for future research (Mele & Magazzino, 2021).

Evidences confirming the impact of COVID-19 on economic indicators offer a valuable information bucket for the policymakers. Governmental decisions to contain and mitigate the pandemic automatically impact the financial scenario of a nation, and frameworks like ours which provide estimates of the magnitude and duration of those impacts allow well-informed and planned implementations of such decisions. The dynamic nature of the models allows them to be updated with new incoming data, thus providing a continuous stream of literate guidance for policymaking.

# APPENDICES

## Appendix 1: Contribution of BRICS Nations in GDP, Population and Export.

| Year | Countries | GDP | Population | Export |
|---|---|---|---|---|
| 1990 | Brazil | 461951782000 | 149003223 | 35170000000 |
| 1990 | Russia | 516814274021.956 | 147969407 | - |
| 1990 | India | 320979026419.633 | 873277798 | 22911050152 |
| 1990 | China | 360857912565.966 | 1135185000 | 57374000000 |
| 1990 | South Africa | 115552349035.441 | 36800509 | 27160238991 |
| 1990 | BRICS | 1776155344043 | 2342235937 | 142615289143.381 |
| 1990 | World | 22626369123313.3 | 5280076284 | 4327825412692.12 |
| 2000 | Brazil | 655420645476.906 | 174790340 | 63716723041 |
| 2000 | Russia | 259710142196.943 | 146596869 | 110520310000 |
| 2000 | India | 468394937262.37 | 1056575549 | 59931697988 |
| 2000 | China | 1211346869605.24 | 1262645000 | 190039384000 |
| 2000 | South Africa | 136361298082.061 | 44967708 | 36995346355 |
| 2000 | BRICS | 2731233892623.52 | 2685575466 | 461203461383.663 |
| 2000 | World | 33618616210474.6 | 6114332536 | 7946637886057.78 |
| 2010 | Brazil | 2208871646202.82 | 195713635 | 231995637790.95 |
| 2010 | Russia | 2208871646202.82 | 142849468 | 441833180000 |
| 2010 | India | 1675615335600.56 | 1234281170 | 348035371769.33 |
| 2010 | China | 6087164527421.24 | 1337705000 | 1603944171443.28 |
| 2010 | South Africa | 375349442837.24 | 51216964 | 107735282477.838 |
| 2010 | BRICS | 11871918420503.9 | 2961766237 | 2733543643481.4 |
| 2010 | World | 66113119131563.3 | 6921871614 | 18910070468396 |
| 2019 | Brazil | 1839758040765.62 | 211049527 | 259792130706.455 |
| 2019 | Russia | 1699876578871.35 | 144373535 | 481492690000 |
| 2019 | India | 2875142314811.85 | 1366417754 | 545706420827.351 |
| 2019 | China | 14342902842915.9 | 1397715000 | 2643376928734.95 |
| 2019 | South Africa | 351431649241.439 | 58558270 | 104845841064.64 |
| 2019 | BRICS | 21109111426606.2 | 3178114086 | 4035214011333.4 |
| 2019 | World | 87697518999809.1 | 7673533972 | 24828684436073 |

**Appendix 2: Percentage Share of BRICS Nations in GDP, Population and Export for 2010 and 2019.**

| Year | Countries | GDP | Population | Export |
|---|---|---|---|---|
| 1990 | Brazil | 2.041652284 | 2.821989967 | 0.812648308 |
| | Russia | 2.284123764 | 2.802410402 | - |
| | India | 1.418605984 | 16.53911328 | 0.529389427 |
| | China | 1.594855589 | 21.49940529 | 1.325700428 |
| | South Africa | 0.51069771 | 0.696969267 | 0.627572427 |
| | BRICS | 7.849935331 | 44.35988821 | 3.29531059 |
| | World | 100 | 100 | 100 |
| 2000 | Brazil | 1.949576513 | 2.858698623 | 0.801807305 |
| | Russia | 0.772518835 | 2.397593983 | 1.39078075 |
| | India | 1.39326061 | 17.28030889 | 0.754176783 |
| | China | 3.603202648 | 20.65057784 | 2.391443862 |
| | South Africa | 0.405612466 | 0.735447536 | 0.465547152 |
| | BRICS | 8.124171071 | 43.92262688 | 5.803755852 |
| | World | 100 | 100 | 100 |
| 2010 | Brazil | 3.34104891 | 2.82746699 | 1.226836453 |
| | Russia | 2.306527794 | 2.063740502 | 2.33649674 |
| | India | 2.534467225 | 17.83161028 | 1.840476334 |
| | China | 9.207196102 | 19.32577018 | 8.481957664 |
| | South Africa | 2.616969849 | 3.66433529 | 4.075668563 |
| | BRICS | 17.95697825 | 42.78851736 | 14.45549158 |
| | World | 100 | 100 | 100 |
| 2019 | Brazil | 2.097845027 | 2.750356326 | 1.046338687 |
| | Russia | 1.938340558 | 1.881447786 | 1.939259775 |
| | India | 3.27847623 | 17.80689001 | 2.197886973 |
| | China | 16.35496991 | 18.21474962 | 10.64646391 |
| | South Africa | 0.400731575 | 0.763119968 | 0.42227707 |
| | BRICS | 24.0703633 | 41.41656371 | 16.25222642 |
| | World | 100 | 100 | 100 |